\documentclass[12pt,pre,aps,superscriptaddress]{revtex4}
\usepackage{graphicx}
\usepackage{bm}
\newcommand{\beq}[2]{\begin{equation}#1\label{#2}\end{equation}}
\newcommand{\ceq}[1]{(\ref{#1})}

\newfont{\mbld}{cmbx10 scaled 800}
\newfont{\cab}{cmsy10 scaled 1200}
\newfont{\scab}{cmsy10 scaled 1000}
\newfont{\bcall}{cmbsy10 scaled 1200}

\begin{document}
\title{Applications of a generalization of the nonlinear sigma model
  with $O(d)$ group of symmetry to 
the dynamics of a constrained chain
}
\author{Franco Ferrari}
\email{ferrari@univ.szczecin.pl}
\author{Jaros{\l}aw Paturej}\email{jpaturej@univ.szczecin.pl}
\affiliation{Institute of Physics and CASA*, University of Szczecin,
  ul. Wielkopolska 15, 70-451 Szczecin, Poland}
\author{Thomas A. Vilgis}
\email{vilgis@mpip-mainz.mpg.de} \affiliation{Max Planck Institute
for Polymer Research, 10
  Ackermannweg, 55128 Mainz, Germany}

\begin{abstract}
Subject of this work are the applications 
of a field theoretical model, called here
generalized nonlinear sigma model
or simply GNL$\sigma$M,
to the dynamics
of a chain subjected to constraints. 
 Chains with similar properties and constraints
have been discussed in a seminal paper of Edwards and Goodyear using an
approach based on the Langevin equation.

The GNL$\sigma$M has been proposed in a previous publication
in order to describe the dynamics of a two dimensional
chain.
In this paper the model is extended to $d$ dimensions and
a bending energy term is added to its action.
As an application, two observables are computed in
the case of a very 
stiff chain.
The first observable is
the dynamical form factor of a ring shaped chain.
The second observable is a straightforward generalization to dynamics
of the static form factor. This
observable is relevant in order to estimate
the average distance between two arbitrary points of the chain.

Finally, a variant of the GNL$\sigma$M is
presented,
in which the topological conditions which constrain the motion of two
 linked chains are imposed with the help of the Gauss linking invariant.
\end{abstract}
\maketitle
\section{Introduction}\label{sec:intro}
Subject of this work is the dynamics of a chain obtained by taking the
continuous limit of a freely jointed chain consisting of $N-1$
segments of  length $a$ and $N$ beads of mass $m$ attached at the
joints between two consecutive segments.
This problem has been addressed in the seminal paper of 
\cite{EdwGoo} using an approach based on the Langevin
equation \footnote{See also the pearl-necklace model proposed in
  \cite{kraemers} and the Verdier--Stockmayer model \cite{VeSt}.}. 
It was shown in \cite{EdwGoo} that the condition of fixed
length segments becomes in the continuous limit a constraint which is
similar to that of  incompressible fluids in
hydrodynamics. The authors of Ref.~\cite{EdwGoo} have also described
several interesting regimes in which their model of a constrained
chain can be applied, like for instance an isolated cold chain or a
hot polymer in the vapor phase. 
The statistical mechanics of a freely jointed
chain in the 
continuous limit has been later investigated 
exploiting different methods, see for example \cite{grosberg,mazars}.
Interesting related results
may be found also in Ref.~\cite{a-e}.
Up to recent times, however, most of the developments
in the dynamics of a chain with rigid constraints have
 been confined to numerical
simulations, see for example Refs.~\cite{ottinger,muthukumar,adland}. 

To overcome at least in part
the complications of the dynamical case, it has been proposed in
Ref.~\cite{FePaVi}  a path integral 
framework for the dynamics of the constrained chain
discussed in  \cite{EdwGoo}. The resulting model, which describes the
fluctuations of a two dimensional chain, is a generalization of the
$O(2)$ nonlinear sigma model. For this reason, it has been called
generalized 
nonlinear sigma model or simply GNL$\sigma$M. 
The  relation of the GNL$\sigma$M with
the Rouse model \cite{rouse} has been studied in details in
Ref.~\cite{FePaVi}. 
A difference between the two models concerns
the scales of time and length at which the chain is observed.
In the Rouse model only the long time-scale behavior of
the chain is considered \cite{DoiEdwards}.
On the other side, the GNL$\sigma$M takes into account the short time-scale
behavior and the finest details of the chain. These facts make the
GNL$\sigma$M suitable to study the response of a chain to mechanical
stresses in micromanipulations, for instance when it is stretched
under a constant 
force \cite{markosiggia,kroy}. Indeed, some experiments point out that the
freely jointed chain model is able to capture the behavior of DNA in
the limit of low applied forces \cite{bustamante}.

The GNL$\sigma$M does not take into account the hydrodynamic
and self-avoiding interactions. The lack of hydrodynamic interactions
limits  
its validity to the cases in which the motion of the beads
is slow. This happens for
instance when the viscosity of the fluid is 
large or the temperature is low.
The conformations of the chain change slowly also in the presence of
stiffness. The treatment of chain stiffness, a feature
which was missing in the formulation of the GNL$\sigma$M of
Ref.~\cite{FePaVi}, will be included in this work. 
A concrete application of 
the GNL$\sigma$M  could be
polymers in a very dilute solutions at the so-called
$\Theta$ point, in which self--avoiding
interactions play no role. The assumption that the chain is phantom,
i.~e. it can 
cross itself, is however dangerous at the $\Theta$ point because in
that case it is very likely that the chain is knotted \cite{deGennes}
and one should take into account the resulting
topological constraints.
In general, the fixing of
constraints in (stochastic) 
dynamics requires some mathematical effort \cite{DoiEdwards,ottinger,
  arti, zinn}. 
The field theoretical
formulation of the chain dynamics provided by the GNL$\sigma$M has the
advantage 
that it is  relatively easy to add further constraints, like for
instance those which are necessary to impose topological conditions in
the case of ring-shaped chains.

The main goal of this work is the development of possible
applications of the 
GNL$\sigma$M model.
The most important result is indeed
the calculation of 
 the expectation values of two
observables in a semiclassical approximation, which is valid if the
changes in the chain conformation due to the fluctuations are small.
This may happen when the chain is relatively rigid or in the following
two cases: The temperature is low or the chain is moving in a very
viscous solution. All these situations are compatible with the
conditions of validity of the GNL$\sigma$M mentioned before.
The first observable which we consider is the dynamical form factor of
the chain 
\cite{DoiEdwards}. The second observable is a straightforward
generalization to dynamics of the static form factor.
It is shown that this observable is related to the average
distance between two points of the chain. 
The calculation of both observables is
complicated by the presence of 
ultraviolet divergences,
which are regulated with the help of
the zeta function regularization \cite{hawking}.
Let us note that divergences of this kind do not appear
in analogous
computations of the dynamical  form factor performed using the Rouse
model 
\cite{formfactcalc}.

Another purpose of the present work
is to improve the formulation of the GNL$\sigma$M given in
\cite{FePaVi}, making it more 
suitable for concrete applications.
For this reason, we consider here
the dynamics of a chain
in $d$ spatial dimensions. 
This case
leads to a GNL$\sigma$M with
$O(d)$
group of 
symmetry, which is a straightforward generalization
of the two dimensional model already discussed in \cite{FePaVi}.
With respect to Ref.~\cite{FePaVi},
we
have also 
included in our approach the bending energy of the
chain.
 In order to make the description of the chain dynamics closer to
 realistic situations, a method to take into account
the topological entanglement of two closed chains is proposed. The
topological constraints are imposed using the Gauss linking invariant.
Unfortunately, it is not possible to apply 
to dynamics
in a straightforward way
the strategy based on Chern-Simons field theory
which is used in the statistical mechanics of
polymers, see for example
Refs.~\cite{edwa,vilbre,vilkol,FeKlLa,kleinertpi,otto}.  
The main difference from statics is that in dynamics
one has  to take
into account the motion in time of the chain. This implies that,
rather
than with the one dimensional trajectory of the chain,
one
has to deal with the two 
dimensional surface that the chain spans during its motion.
To cope with this situation, we have generalized the multi-component
Chern-Simons field 
theory of statistical mechanics to four dimensions.
Mathematically, it is not possible to do that while keeping the
topological invariance of the theory with respect to diffeomorphisms
which depend both on time and on the spatial dimensions.
However, the condition
of invariance under diffeomorphism depending on time is not
strictly
necessary in the case of a non-relativistic chain and has been relaxed.

The presented results are organized as follows.
In Section II  the problem of the dynamics of a chain in $d$
dimensions is mapped into an $O(d)$ GNL$\sigma$M.
The generating functional of the correlation functions of the bond
vectors is expressed in the path integral form.
 In Section III the background field method is applied to
the computation of the generating functional. Particular care is dedicated
to the boundary conditions imposed on the fields to allow the freedom
of performing integrations by part in the action without producing
unwanted and cumbersome boundary terms.
The action of the GNL$\sigma$M is modified in order to take into
account the bending energy of the 
chain.
In Section IV the dynamical form factor and another related
observable are computed.
In Section V a model of two entangled chain is presented.
Finally, our Conclusions are drawn in Section VI.

\section{Formulation of the model} \label{sec:form}

In this section a path integral formulation of the dynamics of a
freely jointed chain of length $L$ is provided. The chain is
regarded as a set of $N$ beads connected together by $N-1$ segments
of fixed length $a$. In addition, $N$, $L$ and $a$ satisfy the
relation $L=Na$. Denoting with $\mathbf R_n(t)$, $n=1,\ldots,N$ the
positions of $N$ beads, it is possible to describe the
fluctuations of the chain as a random walk of the beads constrained by
the conditions:
\beq{
\left|
\mathbf R_n(t) - \mathbf R_{n-1}(t)
\right|^2=a^2\qquad\qquad n=2,\ldots,N
}{discons}
These conditions are required by the fact that the length of the $N-1$
segments connecting the beads is 
equal to $a$.
We also demand that at the initial and final instants $t=0$ and
$t=t_f$ the $n$-th bead is located 
respectively at the positions $\mathbf R_n(0)=\mathbf R_{0,n}$ and
$\mathbf R_n(t_f)=\mathbf R_{f,n}$. 
At this point, following Ref.~\cite{FePaVi}, we introduce the
probability function $\psi_N$ which 
measures the probability that 
the chain after a given time $t_f$ passes from an initial configuration
$\mathbf R_{0,n}$ to a 
final configuration $\mathbf R_{f,n}$.
Using an approach which is widespread in the statistical mechanics of
polymers subjected to constraints,
we define $\psi_N$ as follows:
\begin{eqnarray}
\psi_N&=&\int_{\mathbf R_1(t_f)=\mathbf R_{f,1}\atop
\mathbf R_1(0)=\mathbf R_{0,1}}{\cal D}\mathbf R_1(t)\ldots
\int_{\mathbf R_N(t_f)=\mathbf R_{f,n}\atop
\mathbf R_N(0)=\mathbf R_{0,n}}{\cal D}\mathbf R_N(t)
\exp{\left\lbrace
-\sum_{n=1}^{N}\int_0^{t_f}dt
\frac{\dot\mathbf R_n^2(t)}{4D}
\right\rbrace}
\nonumber\\
&\!\!\!\!\!\!\!\!&\!\!\!\!\!\!\times
\prod^N_{n=2}\delta\left(
\frac{\left|\mathbf R_n(t)-\mathbf R_{n-1}(t)\right|^2}{a^2}-1
\right)
\label{piford}
\end{eqnarray}
where $D$ denotes the diffusion constant.

The path integral \ceq{piford} describes the random walks of the $N$
beads composing the chain. The 
insertion of the Dirac delta functions is needed to enforce the
conditions \ceq{discons}, which describe the rigid constraints 
due to the non extensibility of the individual segments. 
We remember that the diffusion constant $D$ satisfies the relation
$D=\mu k_BT$, where $\mu$ is the 
mobility of a bead, $k_B$ is the Boltzmann constant and $T$ is the
temperature. Moreover, $\mu=\frac \tau m$, 
where $m$ is the mass of the bead and $\tau$ is the relaxation time
which characterizes the ratio of the 
decay of the drift velocity of the beads.
Supposing that the total mass of the chain is $M$, we have of course
that $m=\frac MN=\frac ML a$. 
Thus, Eq.~\ceq{piford} becomes:
\begin{eqnarray}
\psi_N&=&\left[\prod^N_{n=1}\int_{\mathbf R_n(t_f)=\mathbf R_{f,n}\atop
\mathbf R_n(0)=\mathbf R_{0,n}}{\cal D}\mathbf R_n(t)\right]
\exp{\left\lbrace
-\frac{M}{4k_BT\tau L}\sum_{n=1}^{N}a\int_0^{t_f}dt
\dot\mathbf R_n^2(t)
\right\rbrace}
\nonumber\\
&\!\!\!\!\!\!\!\!&\!\!\!\!\!\!\times
\prod^N_{n=2}\delta\left(
\frac{\left|\mathbf R_n(t)-\mathbf R_{n-1}(t)\right|^2}{a^2}-1
\right)
\label{pipsin}
\end{eqnarray}
The limit $N\longrightarrow\infty$, $a\longrightarrow 0$ in which the
 continuous chain is rigorously recovered 
 has been
already discussed in Ref.~\cite{FePaVi} in the two dimensional
 case. The extension to $d$ dimensions is  
straightforward. Basically, the continuous limit consists in the
following replacements of the basic ingredients appearing in the
path integral of Eq.~\ceq{pipsin}:
\begin{eqnarray}
\prod_{n=1}^N\int_{\mathbf R_n(t_f)=\mathbf R_{f,n}\atop
\mathbf R_n(0)=\mathbf R_{0,n}}{\cal D}\mathbf
R_n(t)&\longrightarrow&\int{\cal{D}}\mathbf R(t,s) 
\nonumber\\
\sum_{n=1}^{N}a\int_0^{t_f}dt\dot\mathbf
R_n^2(t)&\longrightarrow&\int_0^{t_f}dt\int_0^Lds\dot\mathbf R^2(t,s) 
\nonumber\\
\prod_{n=2}^N\delta\left(
\frac{\left|\mathbf R_n(t)-\mathbf R_{n-1}(t)\right|^2}{a^2}-1
\right)&\longrightarrow& \delta(\mathbf R^{\prime 2}(t,s)-1)
\label{ddfs}
\\
\mathbf R_{f,n}&\longrightarrow&\mathbf R_f(s)
\nonumber\\
\mathbf R_{0,n}&\longrightarrow&\mathbf R_0(s)\nonumber
\end{eqnarray}
where $s$ is the arc-length of the chain and $0\leq s\leq L$. We have
also introduced the notation 
$\mathbf R^{\prime}\equiv \frac{\partial\mathbf R}{\partial s}$.
Applying Eqs.~\ceq{ddfs} to Eq.~\ceq{pipsin}, the
probability function $\psi_N$ becomes: 
\begin{eqnarray}
\Psi(\mathbf R_f(s), \mathbf R_0(s))&=&
\int_{\mathbf R(t_f,s)=\mathbf R_f(s)\atop
\mathbf R(0,s)=\mathbf R_0(s)}
{\cal{D}}\mathbf R(t,s){\cal {D}}\lambda(t,s)
\exp{\left\lbrace -c\int^{t_f}_0dt\int^L_0ds\dot \mathbf R^2
\right\rbrace}
\nonumber\\
&\!\!\!\!\!\!\!\!&\!\!\!\!\!\!\times
\exp{\left\lbrace i\int^{t_f}_0dt\int^L_0ds\lambda(\mathbf R^{\prime 2}-1)
\right\rbrace}
\label{probfunct}
\end{eqnarray}
with $c=\frac{M}{4k_BT\tau L}$. In the above equation the Lagrange
multiplier $\lambda=\lambda(t,s)$ 
has been introduced in order to represent conveniently  the functional
Dirac delta function appearing 
in the right hand side of Eq.~\ceq{ddfs}.

Formally, the path integral in the right hand side of
Eq.~\ceq{probfunct} resembles the partition 
function of a  quantum mechanical chain 
with constant density mass $\frac M L$ after
the analytical continuation to purely imaginary times:
\beq{S_0=\frac M{2L}\int_0^{t_f}dt\int_0^Lds\dot\mathbf R^2
}{szeroaction}
To stress the close analogy with quantum mechanics, we remark that
in Eq.~\ceq{probfunct}
the action $S_0$ is multiplied by the inverse of the
factor $\kappa=2k_BT\tau$. It is known that $\kappa$ plays 
in the Brownian motion 
the same role of the Planck constant, due to the
well known 
duality 
between quantum mechanics and Brownian motion \cite{rice}.
One may show that the action $S_0$ originates from the continuous
limit of the kinetic energy of a free 
chain, see Ref.~\cite{FePaVi} in the two dimensional case and
Ref.~\cite{FePaVipreprint} in three dimensions. As we see from
Eq.~\ceq{probfunct}, the presence of rigid constraints
is responsible  for 
the appearance besides the action $S_0$ of an additional nonlinear term 
given by
\beq{S_1=-i\int_0^{t_f}dt\int_0^Lds\lambda(\mathbf R^{\prime 2}-1)
}{soneaction}
 The Lagrange
multiplier $\lambda(s,t)$ in $S_1$ closely resembles the
pressure in incompressible
hydrodynamics, as it has been noticed in \cite{EdwGoo}. It expresses the
fact that the segments 
composing the chain have a 
fixed length and  thus they may not be compressed.

To conclude this introductory Section, we specify the set of boundary
conditions 
satisfied by the bond vector 
$\mathbf R(t,s)$ in the probability function \ceq{probfunct}.
First of all, the boundary conditions at the initial and final
instants $0$ and $t_f$ are given by 
\beq{
\mathbf R(0,s)=\mathbf R_0(s) \qquad \mathbf R(t_f,s)=\mathbf R_f(s)
}{bcone}
where $\mathbf R_0(s)$ and $\mathbf R_f(s)$ are static chain
conformations. 
Additionally, it will be convenient to choose boundary conditions with
respect to the arc-length $s$ 
which allow integrations by parts in this variable 
without generating cumbersome boundary terms in the actions
$S_0$ and $S_1$. To this purpose, we will
limit ourselves to the following 
two choices:
\begin{enumerate}
\item periodic boundary conditions in the case of a ring-shaped chain
\beq{\mathbf R(t,s+L)=\mathbf R(t,s)
}{bcrsc}
\item fixed end boundary conditions in the case of an open chain
\beq{
\mathbf R(t,0)=\mathbf r_1 \qquad \mathbf R(t,L)= \mathbf r_2}
{bcocc}
\end{enumerate}
where $\mathbf r_1$ and $\mathbf r_2$ are constant vectors.
For the Lagrange multiplier $\lambda$ one may impose trivial boundary
conditions in time 
\beq{\lambda(0,s)=\lambda(t_f,s)=0
}{lambdaintime}
and boundary conditions analogous to Eq.~\ceq{bcrsc}
in
the case of a ring-shaped chain.
For 
an open chain it is sufficient to require that:
\beq{\lambda(t,0)=0\qquad\qquad\lambda(t,L)=0}{lambdains}

\section{The generating functional $\Psi[J]$ and the bending energy}
\label{sec:gen} 
To the probability function $\Psi(\mathbf R_f(s), \mathbf R_0(s))$
of Eq.~\ceq{probfunct} we associate the following
generating functional $\Psi[J]$:
\beq{
\Psi[J]=
\int_{b.c.}{\cal{D}}\mathbf R{\cal{D}}\lambda
e^{
-\frac{1}{2k_BT\tau}
S_0-S_1
}
e^{
\int_0^{t_f}dt\int_0^L ds \mathbf J\cdot\mathbf R
}
}{genfunct}
where $S_0$ and $S_1$ have been defined in Eqs.~\ceq{szeroaction} and
\ceq{soneaction} and $\mathbf J$ is an external current. The subscript
$b.c.$  
near the integration symbol means that an appropriate set of boundary
conditions among those of 
Eqs.~\ceq{bcone}--\ceq{lambdains} should be chosen.
If $\mathbf J=0$, one obtains back the probability function
$\Psi(\mathbf R_f(s),\mathbf R_0(s))$ of Eq.~\ceq{probfunct}. 
In the right hand side of Eq.~\ceq{genfunct} we perform the following
shift of variables 
\beq{
\mathbf R(t,s)=\mathbf R_b(t,s)+\mathbf R_q(t,s)
}{shiftf}
The field $\mathbf R_q(t,s)$ describes the fluctuations around a fixed
background chain conformation 
$\mathbf R_b(t,s)$.
We require that $\mathbf R_b(t,s)$ satisfies the boundary conditions
\beq{
\left\{
{\begin{array}{*{20}c}
   \mathbf R_b(t_f,s)=\mathbf R_f(s) \\
   \mathbf R_b(0,s)=\mathbf R_0(s)  \\
\end{array}} \right.
}{rbtbc}
with respect to the time $t$.
The boundary conditions in the variable $s$ corresponding to
Eqs.~\ceq{bcrsc} and \ceq{bcocc} are:
\begin{enumerate}
\item ring-shaped chain conformations
\beq{
\mathbf R_b(t,s+L)=\mathbf R_b(s)}{rbsringbc}
\item open chain conformations
\beq{
\mathbf R_b(t,0)=\mathbf r_1 \qquad \mathbf R_b(t,L)=\mathbf r_2
}{rbsopenbc}
\end{enumerate}
Finally, we demand that the background conformation fulfills the
constraint 
\beq{\mathbf R_b^{\prime 2}=1
}{conscondrb}
The fluctuations $\mathbf R_q(t,s)$ obey instead the following
boundary conditions with respect to the time variable $t$: 
\beq{\mathbf R_q(t_f,s)=0 \qquad \mathbf R_q(0,s)=0
}{rqtbc}
In the case of the variable $s$, the analogs of Eqs.~\ceq{rbsringbc}
and \ceq{rbsopenbc} are respectively:
\begin{enumerate}
\item ring-shaped chain conformations
\beq{
\mathbf R_q(t,s+L)=\mathbf R_q(t,s)}{rqsringbc}
\item open chain conformations
\beq{
\mathbf R_q(t,0)=0 \qquad \mathbf R_q(t,L)=0
}{rqsopenbc}
\end{enumerate}
The expression of $\Psi[J]$ in terms of $\mathbf R_b$ and $\mathbf R_q$ is
\beq{
\Psi[J]=e^{-S_b}
\int_{b.c.}{\cal{D}}\mathbf R_q{\cal{D}}\lambda
e^{
-\int_0^{t_f}dt\int_0^L ds \left[c\dot\mathbf R_q^2 - i\lambda(\mathbf
  R^{\prime 2}_q+ 
2\mathbf R_q \cdot \mathbf R_b^{\prime}) 
+ \mathbf R_q \cdot (\mathbf J - 2c\ddot\mathbf R_b)      \right]
}
}{psijejone}
In Eq.~\ceq{psijejone} we have introduced the notation:
\beq{
S_b=\int_0^{t_f}dt\int_0^L ds \left[ c\dot\mathbf R^2_b+\mathbf
  R_b\cdot \mathbf J   \right] 
}{Sb} 
Let us note that it is not necessary that the background field
$\mathbf R_b$  satisfies the
classical equations 
of motion related to the action $\frac{S_0}{4k_BT\tau}+S_1$ appearing in
Eq.~\ceq{genfunct}. 
This would be a severe restriction.
We recall in fact that the compatibility with the constraint 
\ceq{conscondrb}
requires that the solutions of
the classical equations of motion $\mathbf R_{cl}(s)$ are
static chain conformations independent of time
 \cite{FePaVi}.
The main disadvantage of static background conformations
is that with this choice  
the initial and
final conformations of the chain 
in the boundary conditions \ceq{rbtbc}
must be the same, i.e. $\mathbf R_f(s)=\mathbf R_0(s)=\mathbf
R_{cl}(s)$. 
On the other hand, we have seen here that more general  background
fields can be considered. Their only effect is the addition to the
external current $\mathbf J$ of the term $2c\ddot\mathbf R_b$, see
Eq.~\ceq{psijejone}. Of course, for static solutions $\ddot\mathbf
R_b=0$ and this term vanishes identically.

Let's now investigate how it is possible to include in our approach
the stiffness of the 
chain.
To this purpose, it is convenient to
require that 
$\mathbf R_b$ is a static background conformation of the chain.
Under this hypothesis, 
Eq.~\ceq{shiftf} becomes 
$\mathbf R(t,s)=\mathbf R_b(s)+\mathbf R_q(t,s)$. Taking the
derivative with  respect to $s$ of
both members of this equation, one obtains $\mathbf R^{\prime}(t,s)=
\mathbf R_b^{\prime}(s)+\mathbf R^{\prime}_q(t,s)$.
At this point, 
only  the fluctuations are responsible for
the 
time variation of the vector field $\mathbf R'(t,s)$  which is tangent
to the 
chain trajectory for each value of the arc-length $0\le s\le L$.
In fact, the contribution of the background vanishes identically:
$\dot\mathbf R^{\prime}_b(s)=0$. 
Therefore, we may use the module of $\mathbf R^{\prime}_q(t,s)$ as
a measure of how fluctuations 
are effective in bending the chain.
Accordingly, we
modify the generating functional $\Psi[J]$ in
Eq.~\ceq{psijejone} 
adding a bending energy term as follows:
\beq{
\Psi_{\alpha}[J]=e^{-S_b}
\int_{b.c.}{\cal{D}}\mathbf R_q{\cal{D}}\lambda
e^{
-\int_0^{t_f}dt\int_0^L ds \left[c\dot\mathbf R_q^2 +
  \frac{\alpha}{2k_BT\tau}\mathbf R_q^{\prime 2} 
 - i\lambda(\mathbf R^{\prime 2}_q+
2\mathbf R_q'  \cdot \mathbf R_b^{\prime})
+ \mathbf R_q \cdot (\mathbf J - 2c\ddot\mathbf R_b)      \right]
}
}{psijejmod}
The term added to the action of the fields $\mathbf R_q$ and $\lambda$
is:
\beq{S_{\alpha}=\frac{\alpha}{2k_BT\tau}\int_0^{t_f}dt\int_0^Lds\mathbf
R^{\prime 2}_q}{bendact} 
It is easy to realize that the parameter $\alpha$ has the dimension of
an energy per unit of length. 
Thus, $H_{\alpha}(t)=\alpha\int_0^L ds \mathbf R^{\prime 2}_q$ represents the
bending energy measured 
in energy units of $k_BT$
which thermal fluctuations deliver in the unit of time $\tau$ to the system.
To have stiff chains, the fluctuations of the tangent vectors must be
small, i.e. 
\beq{\mathbf R_q^{\prime 2}\ll 1}{stiffnesscond}
 This situation is verified in the
following cases: 
\begin{enumerate}
\item $2k_BT\tau$ is small
\item $\alpha$ is big
\end{enumerate}
The first condition implies that either the temperature is so low
that fluctuations become negligible
or that
the chain is fluctuating in a very viscous environment. In both
situations
one expects that conformational changes due to thermal fluctuations
are small, so that the chain may be considered as rigid.
The interpretation of the second condition is straightforward.
Finally, we notice that it is not possible to substitute naively
$\mathbf 
R'_q$  with $\mathbf R'$
in Eq.~\ceq{bendact} because in this way the bending energy
term would be trivial due to the rigid constraints which require that
$\mathbf 
R^{\prime\,2}=1$.  

\section{Computation of the dynamical form factor and related
  quantities} \label{sec:DSF} 

This Section is devoted to the computation of the average values of
two physically interesting observables. The average $\langle{\cal
  O}\rangle$ of an observable ${\cal O}$ will be evaluated using the
distribution:
\beq{
\int{\cal{D}}\rho(\mathbf R_{q},\lambda)=\int{\cal{D}}\mathbf
R_{q}\int{\cal{D}}\lambda 
e^{-\int_0^{t_f}dt\int^L_0 ds
\left[ c\dot\mathbf R^2_q -i\lambda
(\mathbf R^{\prime\,2}_q+2\mathbf R^{\prime}_q\cdot\mathbf R_b^{\prime})
\right]
}}
{distDSF}
The boundary conditions satisfied by the fields
$\mathbf R_q$ and $\lambda$
are those specified in the previous Section.
Let us note that with this choice of boundary conditions
${\cal{D}}\rho(\mathbf
R_{q},\lambda)$ is exactly the measure appearing in the path integral
of the generating functional of 
Eq.~\ceq{psijejone} as it should be.
In writing
Eq.~\ceq{distDSF} we have required that the 
background field $\mathbf R_b$ satisfies the classical equations of
motion, so that $\ddot\mathbf R_b=0$. 
Without this condition, it is
very difficult to perform analytical calculations, even
in the stiff chain approximation  of Eq.~\ceq{stiffnesscond}.
We have also neglected the stiffness term $S_\alpha$ of
Eq.~\ceq{bendact} and the irrelevant normalization factor $e^{-S_b}$.

First of all, we consider the quantity:
\beq{
\Psi(\bm\xi_1)=\left\langle e^{-\int_0^{t_f}dt \int_0^L ds \mathbf
  R_q(t,s) \cdot \bm\xi_1(t,s)} 
\right\rangle
}{obseone}
where
\beq{
\bm\xi_1(t,s)=i\mathbf k\left[
  \delta(t-t_2)\delta(s-l_2)-\delta(t-t_1)\delta(s-l_1) \right] 
}{currone}
In the above equation $\mathbf k$ is a constant vector and we have assumed
that $0\leq t_1\leq 
t_2\leq t_f$. 
Let us note that the observable \ceq{obseone} is related to the
dynamical form factor, see Ref.~\cite{DoiEdwards} for an introduction
to that 
quantity. As a matter of fact, 
substituting the current \ceq{currone} in  Eq.~\ceq{obseone} and
integrating over $l_1$ and $l_2$ we obtain: 
\beq{
\frac{1}{L^2}\int_0^L dl_1\int_0^{L}dl_2\left[
e^{-\int_0^{t_f}dt\int_0^Lds\mathbf R_b\cdot\bm\xi_1}
\Psi(\bm\xi_1)\right]=
\frac{1}{L^2}\int_0^L dl_1\int_0^{L}dl_2\left\langle e^{i\mathbf
  k\cdot\left( 
\mathbf R(t_1,l_1)-\mathbf R(t_2,l_2)
\right)}
\right\rangle
}{DSF}
where we have added the normalization factor $\frac 1 {L^2}$. The
quantity in the right hand side of the above 
equation is nothing but the dynamical form  factor of the chain.

We are now going to compute the expression of $\Psi(\bm \xi_1)$.
Looking at the integration measure of Eq.~\ceq{distDSF}, it is easy to
realize
that
$\Psi(\bm\xi_1)$ coincides with the generating functional
$\Psi[J]$ in the special case in which 
$\mathbf J$ is the external current $\bm\xi_1$ of Eq.~\ceq{currone}.
As we will see, the presence of Dirac delta functions in $\bm\xi_1$
produces ultraviolet divergences in the expectation value
$\Psi(\bm\xi_1)$ which should be properly regulated.
For simplicity, we will consider here stiff chains.
As explained above, see Eq.~\ceq{stiffnesscond}, this means that the
changes 
due to the fluctuations of the  vectors tangent
to the chain's trajectory are relatively 
small, so that it is possible to neglect in the probability
distribution \ceq{distDSF} the quadratic term in $ R^{\prime}_q$:
\beq{\mathbf R^{\prime 2}_q + 2\mathbf R_q^{\prime}\cdot
\mathbf R^{\prime}_b\sim 
2\mathbf R^{\prime}_q\cdot\mathbf R_b'}{approxcond}

Taking into account Eq.~\ceq{approxcond}, the expression of $\Psi(\bm
\xi_1)$ may be approximated
as follows: 
\beq{\Psi(\bm\xi_1)\sim\int{\cal{D}}\mathbf R_q\int{\cal{D}}\lambda
e^{-\int_0^{t_f}dt\int_0^L ds \left[ c\dot\mathbf R^2_q +\mathbf
    j\cdot \mathbf R_q \right] 
}
}{psiwithj}
where
\beq{\mathbf j=\bm\xi_1 - i\frac{\partial}{\partial s}(\lambda\mathbf
  R_b^{\prime})} {jsmall}
 
After a straightforward gaussian integration over $\mathbf R_q$ in
Eq.~\ceq{psiwithj}, we obtain 
\beq{\Psi(\bm\xi_1)=\int{\cal{D}}\lambda e^{S(\lambda)}}{psixi}
with
\beq{S(\lambda)=\frac 14 \int_0^{t_f}dtdt'\int_0^L ds \mathbf
  j(t,s)G(t,t^{\prime})\mathbf j(t^{\prime},s) 
}{slambda}
In the above formula $G(t,t^{\prime})$ denotes the Green
function satisfying the equation 
\beq{2c\frac{\partial^2G(t,t^{\prime})}{\partial t^2}
=-\delta(t,t^{\prime})
}{eqforG}
The boundary conditions of $G(t,t^{\prime})$
at both initial and final
instants $t=0$ and $t_f=0$  
are the same Dirichlet boundary conditions
 of the fields
$\mathbf R_q$ given in  Eq.~\ceq{rqtbc}.
The Green function $G(t,t^{\prime})$ may be written in closed form as
follows \cite{gfl}: 
\beq{
G(t,t^{\prime})=-\frac{1}{2c}\left[
\frac{t^{\prime}(t-t_f)}{t_f} \theta(t-t^{\prime}) +
\frac{t(t^{\prime}-t_f)}{t_f} \theta(t^{\prime}-t)
\right]
}{Gttprime}
Here $\theta(t)$ is the theta function of Heaviside.
Later on it will be necessary to
evaluate $G(t,t')$ on the line $t=t^{\prime}$.
In this case the right hand side of
Eq.~\ceq{Gttprime} 
becomes proportional to $\theta(0)$, which is not a well defined
quantity.
This is rather a problem of the chosen representation  than an
intrinsic flaw
 of the solution of Eq.~\ceq{Gttprime}.
For this reason, it will be
useful to derive a series representation for
$G(t,t')$, which is regular
when $t=t'$.
To this purpose, we use the definition of a Green function in
terms of its eigenvalues and eigenfunctions:
\beq{
G(t,t^{\prime})=-\sum_{n}\frac{f_n(t)f_n(t^{\prime})}{\lambda_n}
}{Gbyeigen}
where the eigenfunctions $f_n(t)$'s satisfy the equation
\beq{2c\frac{\partial^2}{\partial t^2}f_n(t)=\lambda_n f_n(t)
}{eigeneq}
It is easy to show that
\beq{f_n(t)=\sqrt{\frac{2}{t_f}}\sin{\frac{n\pi(t_f-t)}{t_f}}
}{eifenfunct}
and that
\beq{
\lambda_n=-\frac{2n^2\pi^2c}{t_f^2}
}{eigenval}
where $n>0$ in both Eqs.~\ceq{eifenfunct} and \ceq{eigenval}.
Inserting Eqs.~\ceq{eifenfunct} and \ceq{eigenval} back in \ceq{Gbyeigen}
we obtain:
\beq{G(t,t^{\prime})=
\sum_{n>0}\frac{t_f}{cn^2\pi^2}\sin{\frac{n\pi(t_f-t)}{t_f}}
\sin{\frac{n\pi(t_f-t^{\prime})}{t_f}}
}{Gttprimeser}
Remembering the definition of the current
$\mathbf j$ of Eq.~\ceq{jsmall}, it is easy to realize that the action
$S(\lambda)$ of Eq.~\ceq{slambda} is 
gaussian in the Lagrange multiplier $\lambda$:
\begin{eqnarray}
S(\lambda)&=&\frac 12\int_0^{t_f} dtdt^{\prime} \int_0^L ds
G(t,t^{\prime})\left[ 
\bm\xi_1(t,s)\cdot\bm\xi_1(t^{\prime},s)
+ i\lambda(t,s) \mathbf R^{\prime}_b(s) \cdot \frac{\partial}{\partial
  s}\bm\xi_1(t^{\prime},s)\right. 
\nonumber\\
&\!\!\!\!\!\!\!\!&\!\!\!\!\!\!+
\left.
i\lambda(t^{\prime},s) \mathbf R^{\prime}_b(s) \cdot
\frac{\partial}{\partial s}\bm\xi_1(t,s) 
+ \lambda(t,s) \mathbf R^{\prime}_b(s) \cdot \frac{\partial^2}{\partial s^2}
(\lambda(t^{\prime},s)\mathbf R_b^{\prime}(s))
\right]
\label{actionlambdat}
\end{eqnarray}
At this point we require that the background conformation $\mathbf
R_b(s)$ describes a ring-shaped chain 
placed in a two-dimensional subspace. For instance, we may choose
\beq{
\mathbf R_b(s)=\int_{s_0}^s du \left(   \cos{\varphi(u)},
\sin(\varphi(u)), 0,\ldots,0  \right) 
+\mathbf R_{b,0}
}{rbsub}
$s_0$ is the arc-length of the point $\mathbf
R_{b,0}$ belonging to the background conformation. 
Clearly, the above expression of the background field $\mathbf R_b(s)$
satisfies the constraint 
\ceq{conscondrb} and the periodicity conditions \ceq{rbsringbc} provided the
function $\varphi(u)$ is periodic modulo $2\pi$:
$\varphi(u+L)=\varphi(u) +2k\pi$, $k=0,\pm1,\pm 2,\ldots$.
Using the ansatz
\ceq{rbsub}
it is easy to show that in the action $S(\lambda)$ of
Eq.~\ceq{actionlambdat} 
the last term in the right hand side may be rewritten as
follows:
\begin{eqnarray}
\frac 12\int_0^{t_f} dtdt^{\prime} \int_0^L ds G(t,t^{\prime})
\lambda(t,s) \mathbf R^{\prime}_b(s) \cdot
\frac{\partial^2}{\partial s^2} 
(\lambda(t^{\prime},s)\mathbf R_b^{\prime}(s))=
\nonumber\\
\frac 12\int_0^{t_f} dtdt^{\prime} \int_0^L ds G(t,t^{\prime})
\left[  \lambda(t,s) \left(  \lambda^{\prime\prime}(t^{\prime},s)
-\lambda(t^{\prime},s)\varphi^{\prime 2}(s)     \right)     \right]
\label{lastterm}
\end{eqnarray}
In deriving Eq.~\ceq{lastterm} it has been exploited the fact that
$\mathbf 
R_b'$ is an orthonormal 
vector satisfying the relations $\mathbf R^{\prime 2}_b=1 $ and $\mathbf
R_b'\cdot\mathbf R_b^{\prime\prime}=0$.
Thanks to Eq.~\ceq{lastterm}, the path integral over $\lambda$
\ceq{psixi}  becomes: 
\begin{eqnarray}
\Psi(\bm\xi_1)&=&
\int{\cal{D}}\lambda
e^{
\frac 12\int_0^{t_f} dtdt^{\prime} \int_0^L ds G(t,t^{\prime})
\left[ \bm\xi_1(t,s)\cdot\bm\xi_1(t^{\prime},s)
+i\lambda(t,s)\xi_{1,T}(t^{\prime},s)
+i\lambda(t^{\prime},s)\xi_{1,T}(t,s)\right]}
\nonumber\\
&\!\!\!\!\!\!\!\!&\!\!\!\!\!\!\times
e^{
-\frac 12\int_0^{t_f} dtdt^{\prime} \int_0^L ds G(t,t^{\prime})
 \left[ \lambda^{\prime}(t,s)\lambda^{\prime}(t^{\prime},s)
+\varphi^{\prime 2}(s) \lambda(t,s)\lambda(t^{\prime},s)
\right]
}
\label{psixiT}
\end{eqnarray}
where we have set
\beq{
\xi_{1,T}(t,s)=\mathbf R^{\prime}_b(s)\cdot \bm\xi^{\prime}_1(t,s)}{xioneT}
After a straightforward integration over $\lambda$, we
obtain:
\begin{eqnarray}
\Psi(\bm\xi_1)&=&e^{I_1+I_2}
\label{psifin}
\end{eqnarray}
with 
\beq{I_1\equiv
\frac 12\int_0^{t_f} dtdt^{\prime}\int_0^{L}ds G(t,t^{\prime})
\bm\xi_1(t,s)\bm\xi_1(t^{\prime},s)}{ione}
and
\beq{
I_2\equiv-\frac 12\int_0^{t_f} dtdt^{\prime}\int_0^{L}dsds^{\prime}
  G(t,t^{\prime}) K(s,s^{\prime}) 
\xi_{1,T}(t,s)\xi_{1,T}(t^{\prime},s^{\prime})
}{itwo}
In Eq.~\ceq{itwo}
$K(s,s^{\prime})$ denotes the Green function satisfying the equation
\beq{
\left[\frac{\partial^2}{\partial s^2} - (\varphi^{\prime}(s))^2
  \right] K(s,s^{\prime}) 
=- \delta(s-s^{\prime})
}{eqforK}
At this point, we may proceed our calculation
of $\Psi(\bm\xi_1)$ considering general
background conformations of the form \ceq{rbsub}.
However, one should keep in mind
that, if the function $\varphi$ is too complicated, the solution
of Eq.~\ceq{eqforK} is not known explicitly.
For this reason, we will concentrate here on the particular case in
which 
$\varphi(s)=\frac{2\pi s}{L}$, so that
the background conformation 
$\mathbf R_b(s)$ has the shape of a circle of length $L$:
\beq{
\mathbf R_b(s) = \frac{L}{2\pi}\left( \cos{\frac{2\pi
    s}{L}},\sin{\frac{2\pi s}{L}},0,\ldots,0         \right) 
}{rbscircle}

Let's now evaluate the integrals appearing in the
two exponents in the right hand side 
of Eq.~\ceq{psifin}. 
The explicit calculation of $I_1$ and $I_2$
will be performed in Appendix A. Only
the final results are provided here: 
\beq{I_1
=0 
}{gxioxio}
and
\begin{eqnarray}
I_2&=&
\nonumber\\
&\!\!\!\!\!\!\!\!&\!\!\!\!\!\!\
\sum_{\alpha=1}^2 \left[ \frac \sigma L  g(t_f,t_{\alpha},c) \sum_{i,j=1}^2
\frac{k_ik_j}{2} x^{\prime}_{b,i}(l_{\alpha})
x^{\prime}_{b,j}(l_{\alpha})
+\frac {\sigma L}{4\pi^2} g(t_f,t_{\alpha},c) \sum_{i,j=1}^2
\frac{k_ik_j}{2} x^{\prime\prime}_{b,i}(l_{\alpha}) 
x^{\prime\prime}_{b,j}(l_{\alpha})
\right]
\nonumber\\
&\!\!\!\!\!\!\!\!&\!\!\!\!\!\!\
-\sum_{i,j=1}^2\left\{
\frac 12 k_ik_j \left[G(t_1,t_2) \frac{\partial^2K(l_1,l_2)}{\partial l_1
  \partial l_2} 
x^{\prime}_{b,i}(l_1)x^{\prime}_{b,j}(l_2)
+\frac{\partial K(l_1, l_2)}{\partial l_1}
x^{\prime}_{b,i}(l_1)x^{\prime\prime}_{b,j}(l_2) \right.
\phantom{
\left(
\begin{array}{*{20}c}
   t_1\leftrightarrow t_2  \\
   l_1\leftrightarrow  l_2 \\
\end{array}
\right)
}
\right.
\nonumber\\
&\!\!\!\!\!\!\!\!&\!\!\!\!\!\!\
\left.\left.
+\frac{\partial K(l_1, l_2)}{\partial l_2}
x^{\prime\prime}_{b,i}(l_1)x^{\prime}_{b,j}(l_2) 
+K(l_1,l_2) x^{\prime\prime}_{b,i}(l_1)x^{\prime\prime}_{b,i}(l_2)
+\right]\left(
\begin{array}{*{20}c}
   t_1\leftrightarrow t_2  \\
   l_1\leftrightarrow  l_2 \\
\end{array}
\right)
\right\}
\label{gkxixi}
\end{eqnarray}
In the above equation we have put
\beq{
g(t_f,t_{\alpha},c)=
\sum_{n>0}\frac{t_f}{cn^2\pi^2}\sin^2{\frac{n\pi(t_f-t_{\alpha})}{t_f}}\equiv
G(t_{\alpha},t_{\alpha}) 
}{smallg}
with $\alpha=1,2$
and $\sigma$ being  a constant defined in Eq.~\ceq{aconst}.

The function $K(s,s^{\prime})$ appearing in Eq.~\ceq{gkxixi} is the
Green function of Eq.~\ceq{eqforK}. 
If the background conformation is given by Eq.~\ceq{rbscircle},
$K(s,s^{\prime})$ satisfies the 
relation:
\beq{
\left[
\frac{\partial^2}{\partial s^2} - \frac{4\pi^2}{L^2}
\right]K(s,s^{\prime})=-\delta(s,s^{\prime})
}{ksspcircle}
An explicit expression of the solution of
Eq.~\ceq{ksspcircle} 
in the form of a Fouries series
is given in the Appendix, Eqs.~\ceq{FofK} and \ceq{FoFKtilde}.
 Finally, in Eq.~\ceq{gkxixi}
the components of the background conformation field $\mathbf
R^{\prime}_b$ have been denoted with 
the symbols $x_{b,i}^{\prime}(s)$, $i=1,2$. The particular choice of
$\mathbf 
R_b(s)$ made in Eq.~\ceq{rbscircle} implies 
\beq{
x^{\prime}_{b,1}(s)=\cos{\varphi (s)} \qquad
x^{\prime}_{b,2}(s)=\sin{\varphi (s)} 
}{x1x2}
with $\varphi(s)=\frac{2\pi s}{L}$.
Let us note that the delta functions present in the external current
$\bm\xi_1(t,s)$ of Eq.~\ceq{currone} 
are responsible for the self-interactions of the two points 
on the chain corresponding to the values
of the arc-length $s=l_1$ and $s=l_2$. These self-interactions
introduce infinities in both integrals $I_1$ and $I_2$.
Such infinities have been regulated 
in order to obtain the final result of Eqs.~\ceq{gxioxio} and \ceq{gkxixi}
with the help of a $\zeta$-function 
regularization \cite{hawking}.
At the end it is possible to write:
\beq{\Psi(\bm \xi_1)=e^{I_2}}{psi1final}
where $I_2$ is given in Eq.~\ceq{gkxixi}.

In a way which is analogous  to that used to
calculate the quantity \ceq{obseone} one may
compute also the 
following observable:
\beq{
\Psi(\bm\xi_2)=
\left\langle e^{-\int_0^{t_f}dt\int_0^L ds \bm\xi_2(t,s)\cdot
\mathbf R_q(t,s)}   \right\rangle
}{obsetwo}
where
\beq{
\bm\xi_2(t,s)=\frac{i\mathbf k}{t_2-t_1}\theta(t-t_1)\theta(t_2-t)
\left[  \delta(s-l_1) - \delta(s-l_2)  \right]
}{currtwo}
$\Psi(\bm\xi_2)$ provides a measure of the average distance between
two points of the chain 
over the time $t_2-t_1$.
However, $\Psi(\bm\xi_2)$ may also be used in order to estimate the
distance between two points at any given instant $t_1$.
 As a matter of fact, substituting the expression
of the current \ceq{currtwo} in 
Eq.~\ceq{obsetwo} and taking the limit $t_2\longrightarrow t_1$, it
turns out that:
\beq{
\left.
e^{-\int_0^{t_f}dt\int_0^L ds \bm\xi_2(t,s)\cdot
\mathbf R_q(t,s)}
\Psi(\bm\xi_2)\right|_{t_1=t_2}=
\left\langle e^{i\mathbf k\cdot \left( \mathbf R(t_1,l_2) - \mathbf
  R(t_1,l_1)   \right)}    \right\rangle 
}{psi2}
where $x_i(t,s)$ denotes the $i-$th component of the vector $\mathbf R(t,s)$.
Expanding the exponent in the right hand side of the above equation up
to the second order we obtain
\begin{eqnarray}
\left.\Psi(\bm\xi_2)\right|_{t_1=t_2}&\sim&
\left\langle 1 + i\mathbf k \cdot \left(   \mathbf R(t_1,l_2) - \mathbf R(t_1,l_1)  \right)
-\sum_{i,j=1}^d k_ik_j \left(   x_i(t_1,l_2) - x_i(t_1,l_1)  \right)
\right.
\nonumber\\
&\!\!\!\!\!\!\!\!&\!\!\!\!\!\!\times
\left(   x_j(t_1,l_2) - x_j(t_1,l_1)  \right)
+ \ldots
\Bigg\rangle
\label{psi2sim}
\end{eqnarray}
It is easy to show that, for instance:
\beq{
\left.-\frac{\partial^2}{\partial\mathbf
  k^2}\Psi(\bm\xi_2)\right|_{t_1=t_2 \atop \mathbf k=0} 
= \left\langle |\mathbf R(t_1,l_2)- \mathbf R(t_1,l_1)|^2\right\rangle
}{diffeqforpsi}
confirming the close relation of the observable $\Psi(\bm\xi_2)$ with
the average distance 
of two points of the chain.

The computation of $\Psi(\bm\xi_2)$ can be performed in a way that is
analogous to the calculation 
of $\Psi(\bm\xi_1)$. We report here only the result
\beq{
\left\langle  \Psi(\bm\xi_2)  \right\rangle =
\exp{\left[\sum_{i,j=1}^2  k_ik_j B_{ij} \right]}
}{psixi2fin}
where
\begin{eqnarray}
B_{ij}&=&
\frac A2  x^{\prime}_{b,i}(l_1)x^{\prime}_{b,j}(l_1)\frac{\sigma}{L}
+\frac{A}{2}  x^{\prime}_{b,i}(l_2)x^{\prime}_{b,j}(l_2)\frac{\sigma}{L}
+\frac A2 \frac{\partial^2 K(l_1,l_2)}{\partial l_1\partial l_2}
x^{\prime}_{b,i}(l_1)x^{\prime}_{b,j}(l_2)
\nonumber\\
&\!\!\!\!\!\!\!\!&\!\!\!\!\!\!
+\frac A2 \frac{\partial^2 K(l_2,l_1)}{\partial l_2\partial l_1}
x^{\prime}_{b,i}(l_1)x^{\prime}_{b,j}(l_2)
+A \frac{\partial K(l_1,l_2)}{\partial l_1}
x^{\prime}_{b,i}(l_1)x^{\prime\prime}_{b,j}(l_2) 
\nonumber\\
&\!\!\!\!\!\!\!\!&\!\!\!\!\!\!
+A \frac{\partial K(l_2,l_1)}{\partial l_2}
x^{\prime}_{b,i}(l_2)x^{\prime\prime}_{b,j}(l_1) 
-\frac{\sigma AL}{8\pi^2} \left(
x^{\prime\prime}_{b,i}(l_1)x^{\prime\prime}_{b,j}(l_1) + 
x^{\prime\prime}_{b,i}(l_2)x^{\prime\prime}_{b,j}(l_2)
\right)
\nonumber\\
&\!\!\!\!\!\!\!\!&\!\!\!\!\!\!
+\frac A2 \left(
K(l_1,l_2)x^{\prime\prime}_{b,i}(l_1)x^{\prime\prime}_{b,j}(l_2) 
+K(l_2,l_1)x^{\prime\prime}_{b,i}(l_2)x^{\prime\prime}_{b,j}(l_1)
\right)
\label{Bij}
\end{eqnarray}
and 
\beq{
A=\int_{t_1}^{t_2}\frac{dt}{t_2-t_1}\int_{t_1}^{t_2}\frac{dt'}{t_2-t_1}
G(t,t')
}{adef}

\section{A model of the dynamics of two topologically entangled chains
}\label{sec:topentan}

In this Section we discuss the physically relevant case in which
 $d=3$. 
The  single chain model will be extended
to two chains including topological interactions, which in three space
dimensions become relevant, in particular when the chains are near
the $\Theta-$condition.
 Let us consider two closed chains $C_1$ and $C_2$ of
lengths $L_1$ and $L_2$ respectively. The trajectories of the two
chains are described by the radius vectors $\mathbf R_1(t,s_1)$ and
$\mathbf R_2(t,s_2)$, where $0\le s_1\le L_1$ and $0\le s_2\le L_2$.
The simplest way to impose topological constraints on two closed
trajectories is to use the Gauss linking number $\chi$:
\beq{
\chi(t,C_1,C_2)=\frac{1}{4\pi}\oint_{C_1}d\mathbf R_1\cdot\oint_{C_2}
d\mathbf R_2\times\frac{(\mathbf R_1-\mathbf R_2)}{|\mathbf
  R_1-\mathbf R_2|^3}}{GLN}
If the trajectories of the chains were impenetrable, then $\chi$
would  not depend on time, since it is not possible to change the
topological configuration of a system of knots if their trajectories
are not allowed to cross themselves.
However, since we are not going to introduce repulsive interactions
 between the 
two chains which could prevent their crossing,
we just require that, during the time $t_f$, the
average value of the Gauss linking number is an arbitrary constant
$m$, i.~e.:
\beq{
m=\frac 1{t_f}\int_{0}^{t_f}\chi(t,C_1,C_2)dt}{topconst}
 Our starting point is the
probability function of two free chains:
\beq{
\Psi(C_1,C_2)=\int\prod_{i=1}^2\left[ {\cal{D}}\mathbf R_i
  {\cal{D}}\lambda_i   \right] 
e^{-(S^{(1)}+S^{(2)})}
}{psic1c2}
where, in agreement with Eq.~\ceq{probfunct}, the actions $S^{(1)}$
and $S^{(2)}$ are given by:
\beq{S^{(i)}=\int_0^{t_f} dt \int_0^L ds_i [c\dot\mathbf R^2_i - i
    \lambda_i (\mathbf R^{\prime 2}_i - 1) ] 
\qquad i=1,2 }{Si}
In order to add the topological interactions,
we introduce in the above functional a Dirac
$\delta$-function which imposes the constraint \ceq{topconst}.
In this way we obtain the new probability function:
\begin{eqnarray}
\Psi_m(C_1,C_2)&=&\int\prod_{i=1}^2\left[ {\cal{D}}\mathbf R_i
  {\cal{D}}\lambda_i   \right] 
\delta\left( m - \int_0^{t_f}\frac{dt}{4\pi t_f}
\oint_{C_1}d\mathbf R_1\cdot\oint_{C_2}d\mathbf R_2  \times
\frac{(\mathbf R_1 - \mathbf R_2)}{|\mathbf R_1 - \mathbf R_2|^3}
     \right)
\nonumber\\
&\!\!\!\!\!\!\!\!&\!\!\!\!\!\!\times
e^{-(S^{(1)}+S^{(2)})}
\label{psic1c2const}
\end{eqnarray}
Exploiting the Fourier representation of the Dirac $\delta$-function,
the probability function
$\Psi_m(C_1,C_2)$ takes the form 
\beq{
\Psi_m(C_1,C_2)=\int_{-\infty}^{+\infty}d\Lambda
\Psi_{\Lambda}(C_1,C_2)e^{-im\Lambda}
}{psiFL}
where
\beq{
\Psi_{\Lambda}(C_1,C_2)=
\int\prod_{i=1}^2\left[ {\cal{D}}\mathbf R_i {\cal{D}}\lambda_i   \right]
e^{-(S^{(1)}+S^{(2)})}
e^{ i\Lambda \int_0^{t_f}\frac{dt}{4\pi t_f}
\oint_{C_1}d\mathbf R_1\cdot\oint_{C_2}d\mathbf R_2
\times
\frac{(\mathbf R_1 - \mathbf R_2)}{|\mathbf R_1 - \mathbf R_2|^3
}}
}
{psiLambda}
At this point, after introduction the three dimensional spatial
indices
$\mu,\nu,\rho=1,2,3$, we state the identity:
\begin{eqnarray}
&&
\Lambda\int_0^{t_f}\frac{dt}{4\pi t_f}
\oint_{C_1}d\mathbf R_1\cdot\oint_{C_2}d\mathbf R_2
\times
\frac{(\mathbf R_1 - \mathbf R_2)}{|\mathbf R_1 - \mathbf R_2|^3}
\nonumber\\
&\!\!\!\!\!\!\!\!&\!\!\!\!\!\!
=
\int_{-\infty}^{+\infty} d\eta \int d^3x  \int_{-\infty}^{+\infty}
d\eta^{\prime} \int d^3y 
J_1^{\mu}(\eta,\mathbf x) G_{\mu\nu}(\eta,\eta^{\prime};\mathbf x,
\mathbf y)  J_2^{\nu}(\eta^{\prime},\mathbf y) 
\label{constbycurr}
\end{eqnarray}
In the above equation we have defined the following currents:
\beq{
J^{\mu}_i(\eta,\mathbf x)=\gamma_i \int_0^{t_f}\frac{dt}{t_f}
\delta(\eta-t) 
\int_0^{L_i} ds_i \frac{\partial \mathbf R_i(t,s_i)}{\partial s_i}
\delta^{(3)} (\mathbf x - \mathbf R_i(t,s_i))
}{currentsmui}
with $\gamma_1=\frac 1{2t_f}$ and $\gamma_2=\Lambda$.
$G_{\mu\nu}(\eta,\eta^{\prime};\mathbf x,\mathbf y)$ is the propagator of
the field theory
\begin{equation}
S_{CS}=\frac 1{t_f}\int_{-\infty}^{+\infty} d\eta\int d^3 x
\mathbf A^{(1)}(\eta,\mathbf
x)\cdot(\bm \nabla_{\mathbf x}\times\mathbf A^{(2)}(\eta,\mathbf x))
\label{Scs}
\end{equation}
$\bm \nabla_{\mathbf x}$ being the gradient with respect to the
spatial variable $\mathbf x$. Moreover, the $\mathbf A^{(i)}(\eta,\mathbf
x)'$s, $i=1,2$, are two vector 
fields defined in the Euclidean four dimensional space
$(\eta,\mathbf x)$ and having three spatial components $A^{(i)}_\mu$.
Explicitly, $G_{\mu\nu}(\eta,\eta^{\prime};\mathbf x,\mathbf y)$ is
given by 
\beq{
G_{\mu\nu}(\eta,\eta^{\prime};\mathbf x,\mathbf y)=\frac{t_f}{2\pi}
\varepsilon_{\mu\nu\rho}\frac{(x-y)^{\rho}}{|\mathbf x - \mathbf
  y|^3}\delta(\eta-\eta') 
}{greenCS}
Apparently, $S_{CS}$ is similar to the multi-component
Chern-Simons field theory used
to impose 
the topological constraints in 
the case of a static chain \cite{FeLa}. It differs however from it by
the addition of the 
fourth dimension spanned by the coordinate $\eta$, with
$-\infty<\eta<+\infty$. 
This new coordinate is necessary 
to deal with the time variable $t$ appearing in the dynamical case. 
Using the identity \ceq{constbycurr},
 it is possible to formulate the probability function
$\Psi_{\Lambda}(C_1,C_2)$ of Eq.~\ceq{psiLambda} as a Chern-Simons
 field theory 
\beq{
\Psi_{\Lambda}(C_1,C_2)=\int\prod_{i=1}^2 \left[ {\cal{D}}\mathbf R_i
  {\cal{D}}\lambda_i 
{\cal{D}}\mathbf A^{(i)}\right]
e^{-(iS_{CS} + S^{(1)}   +   S^{(2)})}
e^{-i\sum_{i=1}^2   \int_{-\infty}^{+\infty}d\eta \int d^3x
  J_i^{\mu}(\eta, \mathbf x) 
A_{\mu}^{(i)}(\eta, \mathbf x)}
}{newpsiL}
where the actions $S_{CS}, S^{(1)}, S^{(2)}$ have been defined
respectively defined respectively in 
Eqs.~\ceq{Scs} and \ceq{Si}, while the currents $J^{\mu}_i(\eta,
\mathbf x)$ are given in Eq.~\ceq{currentsmui}. 

\section{Conclusions}
The main goal of this paper was to make the
GNL$\sigma$M of Ref.~\cite{FePaVi} more suitable to describe realistic
systems and to compute the average of concrete physical quantities.
In the introductory Section II, the continuous limit of a freely
jointed chain 
consisting of $N-1$ segments of fixed length $a$ and $N$ beads of mass
$m$ attached at the joints has been discussed in $d-$dimensions.
The final model describing the dynamics of the
continuous chain is a generalized nonlinear sigma model. The
difference from the two dimensional case 
discussed in Ref. \cite{FePaVi}
is that the underlying
symmetry group is $O(d)$ and not $O(2)$. This slight difference
is enough to complicate the calculation of the
generating functional $\Psi[J]$ of Eq.~\ceq{genfunct} even in the 
approximation of Eq.~\ceq{stiffnesscond}.
To obtain analytical results, one is forced to assume that
the background conformations 
$\mathbf R_b$ are lying on a plane as it has been done in
Eq.~\ceq{rbsub}.

In Section III we have introduced in our approach the
notion of chain stiffness. The bending energy term $S_\alpha$ of
Eq.~\ceq{bendact} 
has been added 
 to the action of the GNL$\sigma$M
in Eq.~\ceq{psijejmod}. Let us
note that, due 
to the constraint $\mathbf R_q'^2+2\mathbf R_b'\cdot\mathbf R_q'=0$,
$S_\alpha$ may be treated as a linear term, in which the
fluctuation $\mathbf R_q$ is coupled to an
 external
current proportional to $\mathbf R^{\prime\prime}_b$.
%

The expectation values of two important observables have
been derived 
in Section IV. The first observable $\Psi(\bm\xi_1)$ is the dynamical
form factor. The second observable $\Psi(\bm \xi_2)$ may be related
both 
to the
average distance between two arbitrary points of the chain at a given
time $t_1$ or to the average of that distance over a finite
time interval.
We have seen that the calculation of these observables is complicated
by ultraviolet 
divergences, which have been cured using the zeta function
regularization.
The closed form of $\Psi(\bm\xi_1)$ and  $\Psi(\bm \xi_2)$ in the
approximation \ceq{stiffnesscond} has been presented in
Eqs.~\ceq{gkxixi}--\ceq{psi1final}
 and 
 \ceq{psixi2fin}--\ceq{adef} respectively.

Finally, we have proposed in Section V a way to 
describe the topological relations between two ring--shaped
chains via the Gauss linking number. This result generalizes
to dynamics the treatment of topological constraints presented in the
case of statistical mechanics in \cite{FeLa}.

\section{Acknowledgments} 
This work has been financed
by the Polish Ministry of Science and Higher Education, scientific
project N202 156 
31/2933.  
F. Ferrari gratefully acknowledges also the support of the action
COST~P12 financed by the European Union and the hospitality of
C. Schick at the University of Rostock.
The authors  would like to thank
V. G. Rostiashvili for fruitful discussions.

\begin{appendix}

\section{Proof of Eqs.~\ceq{gxioxio} and \ceq{gkxixi}}
First we  evaluate the integral $I_1$ of Eq.~\ceq{ione}, which we
rewrite here for convenience:
\beq{I_1=
\frac 12 \int_0^{t_f} dtdt^{\prime} \int_0^L ds G(t,t^{\prime})
\bm\xi_1(t,s)\bm\xi_1(t^{\prime},s)
}{I1}
where $\bm\xi_1(t,s)$ is given in Eq.~\ceq{currone}. The only
potentially non-zero 
contributions come from the self-interactions of the two points located
at arc-lengths 
$s=l_1$ and $s^{\prime}=l_2$:
\beq{I_1=
-\frac{\mathbf k^2}{2} \sum_{\alpha=1}^{2} \int_0^{t_f} dtdt^{\prime}
\int_0^L dsds^{\prime} 
G(t,t^{\prime}) \delta(t-t_{\alpha})\delta(t^{\prime}-t_{\alpha})
\delta(s-s^{\prime})\delta(s-l_{\alpha})\delta(s^{\prime}-l_{\alpha})
}{I1withaddsp}
There are in principle other two contributions which are proportional
to $\delta(l_1-l_2)$ and thus vanish identically, since $l_1\ne l_2$.
The time integrations in Eq.~\ceq{I1withaddsp} do not pose particular
problems. Using the
prescription \ceq{Gttprimeser} to evaluate the Green function $G(t,t')$
at coinciding points,
we obtain
\beq{I_1=
-\frac{\mathbf k^2}{2}\sum_{\alpha=1}^2 g(t_f,t_{\alpha},c)  \int_0^L
dsds^{\prime} 
\delta(s-s^{\prime})\delta(s-l_{\alpha})\delta(s^{\prime}-l_{\alpha})
}{I1withoutt}
Here $g(t_f,t_{\alpha},c)$ is the series given in Eq.~\ceq{smallg}.
Unfortunately, the integrals  over the arc-length $s$ and $s^{\prime}$
are divergent and require regularization. 
To this purpose, we first expand the periodic $\delta$-function
$\delta(s)$ in Fourier series :
\beq{
\delta(s)=\sum_{\kappa=-\infty}^{+\infty}\frac{e^{2\pi i \kappa \frac
    sL}}{L} 
}{pdeltainF}
After some calculations, it is possible to show in this way that:
\beq{
\int_0^L dsds^{\prime}
\delta(s-s^{\prime})\delta(s-l_{\alpha})\delta(s^{\prime}-l_{\alpha})=
\frac 1L \sum_{\kappa=-\infty}^{+\infty} 1
}{intdsetc}
Of course, $\sum_{\kappa=-\infty}^{+\infty} 1$  is divergent if left
without treatment. To remove the singularities, we will use the zeta
function regularization. This kind of regularization is based on
the Riemann $\zeta$-function:
\beq{\zeta(s)=\sum_{\kappa=0}^{+\infty} \frac 1{\kappa^s}
}{zeta}
and on the fact that, in the sense of the analytic continuation, one
may write the following formal
identity:
\beq{\zeta(0)=\sum_{\kappa=0}^{+\infty} 1}{zetazero}
On the other side, Eq.~\ceq{zetazero} implies that:
\beq{\sum_{\kappa=-\infty}^{+\infty} 1=2\zeta(0)-1}{zetazerodoubinf}

Applying Eqs.~\ceq{zetazero} and \ceq{zetazerodoubinf}
to Eq.~\ceq{intdsetc} and substituting the
result in the expression of $I_1$ given in \ceq{I1withoutt},
it is easy to realize that the integral $I_1$ becomes:
\beq{
I_1=
-\frac{\mathbf k^2}{2} 
g(t_f,t_\alpha,c)
\sum_{\alpha=1}^{2} \frac 1L (2\zeta(0)-1)
}{I1almost}
After an analytic continuation of the function $\zeta(s)$
 to the point $s=0$, one finds that
$\zeta(0)=\frac 12$. 
Substituting this value of $\zeta(0)$ in Eq.~\ceq{I1almost}, we obtain
\beq{I_1=0
}{I1final}
This completes the proof of
Eq.~\ceq{gxioxio}.
As expected, the self-interactions of the points at $s=l_1$ and
$s=l_2$ with themselves 
do not give any contribution to $\Psi(\bm\xi_1)$. 

 The situation
is more complicated in the case of the second integral $I_2$
of Eq.~\ceq{itwo}:
\beq{
I_2=-\frac 12 \int_0^{t_f} dtdt^{\prime} \int_0^L dsds^{\prime}
G(t,t^{\prime}) K(s,s^{\prime}) 
\xi_{1,T}(t,s) \xi_{1,T}(t^{\prime},s^{\prime})
}{I2}
Using the definition \ceq{xioneT} of the
current $\xi_{1,T}$
one obtains:
\begin{eqnarray}
I_2&=&
\sum_{i,j=1}^2 \sum_{\alpha=1}^2 \frac{k_ik_j}{2} g(t_f,t_{\alpha},c)
\int_0^Ldsds^{\prime} 
\left[
K(s,s^{\prime})x^{\prime\prime}_{b,i}(s)x^{\prime\prime}_{b,j}(s^{\prime})+
\frac{\partial^2 K(s,s^{\prime})}{\partial s\partial s^{\prime}}
x^{\prime}_{b,i}(s)x^{\prime}_{b,j}(s^{\prime}) 
\right.
\nonumber\\
&\!\!\!\!\!\!\!\!&\!\!\!\!\!\!
\left.
+\frac{\partial K(s,s^{\prime})}{\partial s}
x^{\prime}_{b,i}(s)x^{\prime\prime}_{b,j}(s^{\prime}) 
+
\frac{\partial K(s,s^{\prime})}{\partial s^{\prime}}
x^{\prime\prime}_{b,i}(s)x^{\prime}_{b,j}(s^{\prime}) 
\right]\delta(s-l_{\alpha})\delta(s^{\prime}-l_{\alpha})
\nonumber\\
&\!\!\!\!\!\!\!\!&\!\!\!\!\!\!
-\left\lbrace
\sum_{i,j=1}^2  \frac{k_ik_j}{2} G(t_1,t_2) \int_0^L dsds^{\prime}
\left[
K(s,s^{\prime})x^{\prime\prime}_{b,i}(s)x^{\prime\prime}_{b,j}(s^{\prime})+
\frac{\partial^2 K(s,s^{\prime})}{\partial s\partial s^{\prime}}
x^{\prime}_{b,i}(s)x^{\prime}_{b,j}(s^{\prime}) 
\right.\right.
\nonumber\\
&\!\!\!\!\!\!\!\!&\!\!\!\!\!\!
\left.
+\frac{\partial K(s,s^{\prime})}{\partial s}
x^{\prime}_{b,i}(s)x^{\prime\prime}_{b,j}(s^{\prime}) 
+
\frac{\partial K(s,s^{\prime})}{\partial s^{\prime}}
x^{\prime\prime}_{b,i}(s)x^{\prime}_{b,j}(s^{\prime}) 
\right]\delta(s-l_{\alpha})\delta(s^{\prime}-l_{\alpha})
\nonumber\\
&\!\!\!\!\!\!\!\!&\!\!\!\!\!\!
\left.
+
\left(
\begin{array}{*{20}c}
   t_1\leftrightarrow t_2  \\
   l_1\leftrightarrow  l_2 \\
\end{array}
\right)
\right\rbrace
\label{I2bysum}
\end{eqnarray}
To write Eq.~\ceq{I2bysum}, we have made some integrations by parts in
the variables $s$ and $s'$. These are allowed because of the choice of
the boundary condition and of the fact that the current $\bm\xi_1$
vanishes at the boundary: $\bm\xi_1(t,0)=\bm\xi(t,L)=0$.
It is not difficult to show that, for symmetry reasons, $\left.
\frac {\partial K(s,s^{\prime})}{\partial s}\right|_{s=l_{\alpha}\atop
  s^{\prime}=l_{\alpha}} 
=0$.
As a matter of fact, using the Fourier representation of $K(s,s^{\prime})$:
\beq{
K(s,s^{\prime})=\sum_{\kappa=-\infty}^{+\infty} e^{\frac{2\pi i
    \kappa}{L}(s-s^{\prime})} 
\tilde K(\kappa)
}{FofK}
where 
\beq{\tilde K(\kappa)=\frac L{4\pi^2}\frac 1{\kappa^2+1}}{FoFKtilde}
we obtain:
\beq{\left.
\frac {\partial K(s,s^{\prime})}{\partial s}\right|_{s=l_{\alpha}\atop
  s^{\prime}=l_{\alpha}} 
=i\frac 1{2\pi}\sum_{\kappa=-\infty}^{+\infty} \frac
{\kappa}{\kappa^2+1}=0 
}{ks0}
Analogously, one may show that
$\left.\frac {\partial K(s,s^{\prime})}{\partial
  s^{\prime}}\right|_{s=l_{\alpha}\atop s^{\prime}=l_{\alpha}}=0$. 

There is only one term in the expression of $I_2$ which is
divergent and needs regularization. 
This is given by:
\beq{
I_{2,sing}=\sum_{i,j=1}^2\sum_{\alpha=1}^2 \frac {k_ik_j}{2}
g(t_f,t_{\alpha},c) 
\int_0^L dsds^{\prime}
\frac {\partial^2 K(s,s^{\prime})}{\partial s\partial s^{\prime}}
x^{\prime}_{b,i}(s)x^{\prime}_{b,j}(s^{\prime})\delta(s-l_{\alpha})\delta(s^{\prime}
- l_{\alpha}) 
}{firstof2}
Exploiting the Fourier representation
\ceq{FofK}
of $K(s,s^{\prime})$, it turns out that $I_{2,sing}$
may be rewritten as follows:
\beq{
I_{2,sing}=-\sum_{i,j=1}^2\sum_{\alpha=1}^2 \frac {k_ik_j}{2}
g(t_f,t_{\alpha},c)x'_{b,i}(l_\alpha) x'_{b,j}(l_\alpha)
\frac 1L \sum_{\kappa=-\infty}^{+\infty} \frac{\kappa^2}{\kappa^2+1}
}{firstof2bla}
Applying also the identity $\sum_{k=-\infty}^{+\infty}
\frac{\kappa^2}{\kappa^2+1}=\sum_{\kappa=-\infty}^{+\infty} 
\left( 1-\frac{1}{\kappa^2+1}\right)$ and the fact that $\sum_{\kappa
=-\infty}^{+\infty} 1 =0$ as we have previously seen,
we arrive at the final result in which the singularity of $I_{2,sing}$
has been regulated:
\beq{
I_{2,sing}=\frac \sigma L \sum_{i,j=1}^2 \sum_{\alpha=1}^2 \frac
{k_ik_j}2 g(t_f,t_{\alpha},c) 
x^{\prime}_{b,i}(l_\alpha)x^{\prime}_{b,j}(l_\alpha)
}{firstof2fin}
In the above equation we have put:
\beq{
\sigma=\sum_{\kappa=-\infty}^{+\infty} \frac 1{\kappa^2+1}
}{aconst}
Let's now simplify the following term contained in $I_2$:
\beq{
I_{2,0}=\sum_{i,j=1}^2\sum_{\alpha=1}^2 \frac {k_ik_j}{2}
g(t_f,t_{\alpha},c) 
\int_0^L dsds^{\prime}
x^{\prime\prime}_{b,i}(s)x^{\prime\prime}_{b,j}(s^{\prime})
\delta(s-l_{\alpha})\delta(s^{\prime} 
- l_{\alpha}) 
}
{2ndof2}
After the integrations over $s$ and $s^{\prime}$, one obtains from
$I_{2,0}$ an
expression which is proportional 
to the Green function $K(s,s^{\prime})$ computed at coinciding points
$s=s^{\prime}=l_\alpha$.
Exploiting the Fourier representation
\ceq{FofK},
it is possible to check  that $K(l_\alpha,l_\alpha)$ is convergent and
is equal to: 
\beq{
K(l_{\alpha}, l_{\alpha})=\frac{\sigma L}{4\pi^2}}{karrow}
where $\sigma$ is the constant given in Eq.~\ceq{aconst}. 
Thus, we may write:
\beq{
I_{2,0}=
\sum_{i,j=1}^2\sum_{\alpha=1}^2 \frac {k_ik_j}{2}
g(t_f,t_{\alpha},c) \frac{\sigma L}{4\pi^2}
x^{\prime\prime}_{b,i}(l_\alpha)x^{\prime\prime}_{b,j}(l_\alpha) 
}{itwozero}
All the other terms present in $I_2$ are divergenceless.
At the end, remembering
the expressions of the contributions $I_{2,sing}$ and
$I_{2,0}$ to $I_2$ given in 
 Eqs.~\ceq{firstof2fin} and \ceq{itwozero} respectively,
we
obtain the final result:
\begin{eqnarray}
I_2&=&I_{2,sing}+I_{2,0}-
\left\lbrace  \sum_{i,j=1}^2
\frac{k_ik_j}{2} G(t_1,t_2)
\left[
\frac{\partial^2 K(l_1,l_2)}{\partial l_1\partial l_2}
 x^{\prime}_{b,i}(l_{1})
x^{\prime}_{b,j}(l_{2})
+\frac{\partial K(l_1, l_2)}{\partial l_1}
x^{\prime}_{b,i}(l_1)x^{\prime\prime}_{b,j}(l_2) 
\right.\right.
\nonumber\\
&\!\!\!\!\!\!\!\!&\!\!\!\!\!\!\
\left.\left.
+\frac{\partial K(l_1, l_2)}{\partial l_2}
x^{\prime\prime}_{b,i}(l_1)x^{\prime}_{b,j}(l_2) 
+K(l_1,l_2)x^{\prime\prime}_{b,i}(l_1)x^{\prime\prime}_{b,j}(l_2)
+\left(
\begin{array}{*{20}c}
   t_1\leftrightarrow t_2  \\
   l_1\leftrightarrow  l_2 \\
\end{array}
\right)
\right]
\right\rbrace
\label{i2final}
\end{eqnarray}
The above equation coincides exactly with Eq.~\ceq{gkxixi}.
\end{appendix}

\end{document}